\documentclass[10pt]{iopart}

\usepackage{graphicx}
\usepackage{bm}
\usepackage[urlcolor=blue,colorlinks=true,citecolor=blue,linkcolor=blue,pdfstartview={FitH},bookmarks=false]{hyperref}
\usepackage{amssymb}
\usepackage{xcolor}
\usepackage{epstopdf}

\usepackage{cite}

\begin{document}

\title[Multiple phase transitions in Pauli limited iron-based superconductors]
{Multiple phase transitions in Pauli limited iron-based superconductors}

\author{Andrzej Ptok}
\address{Institute of Nuclear Physics, Polish Academy of Sciences, ul. Radzikowskiego 152, PL-31-342 Krak\'{o}w, Poland}
\eads{\mailto{aptok@mmj.pl}}
\vspace{10pt}

\begin{indented}
\item[]February 2014
\end{indented}

\begin{abstract}
Specific heat measurements have been successfully used to probe unconventional superconducting phases in one-band heavy-fermion and organic superconductors. We extend the method to study successive phase transitions in multi-band materials such as iron based superconductors. The signatures are multiple peaks in the specific heat, at low temperatures and high magnetic field, which can lead the experimental verification of unconventional superconducting states with non-zero total momentum. 
\end{abstract}


\vspace{2pc}

\vspace*{2pc}

\noindent (Some figures may appear in colour only in the online journal)

\maketitle
%
%

\section{Introduction}

Ideal diamagnetism is one of most striking properties of superconductors, which is manifested by the {\it Meissner effect}, the complete expulsion of a magnetic field from the volume of a superconductor. Conversely, magnetic fields with relatively large value can destroy superconductivity. In most real (II type) superconductors it may take place in two ways -- by orbital or paramagnetic effects. The orbital pair-breaking is connected with the rise of the Abrikosov vortex state in superconductors, while paramagnetic pair-breaking originates from the Zeeman splitting of electronic energy levels. Both effects determine the {\it upper critical magnetic field}, in which the relative importance of the orbital and paramagnetic effects in the suppression of the superconductivity is described by the Maki parameter $\alpha = \sqrt{2} H_{c2}^{orb} / H_{c2}^{P}$~\cite{maki.66}, the ratio of the critical magnetic fields at zero temperature $H_{c2}^{orb}$ and $H_{c2}^{P}$, derived from orbital and diamagnetic effects respectively.

In most superconductors, the orbital pair-breaking effects are more disruptive than the diamagnetic ones, thus $H_{c2}^{P}$ is usually larger than $H_{c2}^{orb}$ ($\alpha \ll 1$) and superconductivity disappears when vortex cores begin to overlap. When $\alpha \geq 1$, superconductivity is destroyed by the Zeeman effect (Pauli paramagnetism), and systems exhibiting this property are called {\it Pauli limited materials}.

In absence of an external magnetic field, superconductors are found in the  Bardeen-Cooper-Shrieffer (BCS) state~\cite{bcs1,bcs2}, where superconductivity is formed by Cooper pairs with total momentum equal zero. However, in Pauli-limited superconductors, the external field can lead to interesting phenomena (near or) above the critical magnetic field $H_{c2}^{P}$, such as the transition to the Fulde-Ferrell-Larkin-Ovchinnikov (FFLO) phase~\cite{FF,LO}. In contrast to the BCS state, now the Cooper pairs are formed between the spin-up and spin-down sheets of the split Fermi surface with non-zero total momentum. Moreover this phase exhibits a spatially oscillating superconducting order parameter (SOP) in real space and spin polarization.

Although the FFLO phase was already theoretically described in the 1960s, the experimental search is still ongoing and riddled with difficulties. They come as a consequence of the physical properties of this inhomogeneous phase: it can occur only in Pauli limited superconductors at low temperature and high magnetic field (LTHM) regime. Only in the last decade systems have appeared in which we expect an experimental verification of this phase, such as heavy-fermion superconductors~\cite{matsuda.shimahara.07} (e.g. CeCoIn$_{5}$~\cite{bianchi.movshovich.03,bianchi.movshovich.02,kumagai.saitoh.06,mitrovic.horvatic.06,miclea.nicklas.06}) or organic superconductors (e.g. $\beta''$-(ET)$_{2}$SF$_{5}$CH$_{2}$CF$_{2}$SO$_{3}$~\cite{cho.smith.09}, $\lambda$-(BETS)$_{2}$FeCl$_{4}$~\cite{uji.kodama.13}, $\kappa$-(BEDT-TTFS)$_{2}$Cu(NCS)$_{2}$~\cite{lortz.wang.07,bergk.demuer.11,agosta.jin.12,mayaffre.kramer.14,tsuchiya.yamada.15}).  These chemical compounds are often modeled as effectively {\it quasi}-2D one-band systems. However in 2013, D. Zocco {\it et al.} in Ref.~\cite{zocco.grube.13} reported possibly the FFLO phase in the multi-band iron-based superconductor KFe$_{2}$As$_{2}$. Although it should be noted that hints about the FFLO phase in pnictides had been reported in other experimental~\cite{fuchs.drechsler.08,fuchs.drechsler.09,khim.lee.11,cho.kim.11,tarantini.gurevich.11,terashima.kihou.13} and theoretical~\cite{ptok.crivelli.13} works previously.

In experiments on the ordered phases (such as superconducting or magnetic), measurements of the anomalies in the specific heat $C$ are one of the most sensitive tools~\cite{stewart.11}. Using this method we can study phase transitions and their type, or the nodes of the gap function in the superconducting state~\cite{park.bauer.08,jang.vorontsov.11,gofryk.sefat.12}. In this sense peaks in the specific heat reveal information about the phase transition. Moreover, a narrow peak in $C (T)$ is associated with a first order transition, while the $\lambda$-shape behavior is typical for the second order transition~\cite{kapcia.robaszkiewicz.12,kapcia.robaszkiewicz.13}. This method can be used to find phase transition e.g. in multiband superconductors like IBSC \cite{hardy.bohmer.13,popovich.boris.14,griunenko.efremov.14,johnston.abdelhafirez.14} or MgB$_{2}$~\cite{dolgov.kremer.05,golubov.kortus.02}. Specific heat measurements have been also successfully used to investigate the FFLO phase in heavy-fermion systems~\cite{bianchi.movshovich.02,bianchi.movshovich.03,miclea.nicklas.06} and organic superconductors~\cite{lortz.wang.07}, where the peaks, deep in the superconducting state in the LTHM regime, have been interpreted as phase transitions from BCS to FFLO state.

Because the experimental evidence to confirm the FFLO phase in iron-based superconductors (IBSC) is still needed, the description of the theoretical properties of this phase is paramount. In this paper we discuss experimental consequences multiple phase transitions induced by the FFLO phase in Pauli limited multi-band iron-based superconductors at low temperature and high magnetic field.

\section{Model and theoretical background}

IBSC are chemical compounds possessing layered structure, with a characteristic Fermi surface (FS). In order to explain the FS features in IBSC, various two-orbital~\cite{raghu.qi.08}, three-orbital~\cite{daghofer.nicholson.10,daghofer.nicholson.12} and five-orbital~\cite{kuroki.onari.08,graser.maier.09} tight binding models have been proposed. The abundance of available models is induced by the relatively complicated band structure of IBSC materials, which is strongly dependent on the chemical doping~\cite{kordyuk.zabolotnyy.13}. To realistically describe e.g. FeAs layers, five 3d-orbitals of iron ions need to be retained in the model. However, the band structure calculations suggest the importance of itinerant electrons from the d$_{xz}$ and d$_{yz}$ orbitals. Influential in the formation of the FS are also the d$_{xy}$ orbitals. For this reason,  in our calculations we use for simplicity the three-band model proposed by M. Daghofer {\it et al.}~\cite{daghofer.nicholson.10,daghofer.nicholson.12}.

In general, the momentum-dependent tight-binding non-interacting Hamiltonian of the multi-orbital IBSC in orbital space is be given by:
\begin{equation}
H_{0} = \sum_{\alpha\beta} \sum_{{\bm k}\sigma} \left ( T^{\alpha\beta}_{\bm k} - \left( \mu + \sigma h \right) \delta_{\alpha\beta} \right) c_{\alpha{\bm k}\sigma}^{\dagger} c_{\beta{\bm k}\sigma} ,
\end{equation}
where $c_{\alpha{\bm k}\sigma}$ ($c_{\alpha{\bm k}\sigma}^{\dagger}$) annihilates (creates) an electron with momentum ${\bm k}$ and spin $\sigma$ in the orbital $\alpha$. The hopping matrix elements $T_{\bm k}^{\alpha\beta}$ correspond to the kinetic energy of a particle with momentum ${\bm k}$ changing the orbital from $\beta$ to $\alpha$, they are given by the effective tight-binging model of the two-dimensional FeAs planes in the selected model. We use the model of IBSC proposed by M. Daghofer {\it et al.} in Ref.~\cite{daghofer.nicholson.10} and improved in Ref.~\cite{daghofer.nicholson.12}. Beyond the $d_{xz}$ and $d_{yz}$ orbitals, the model also accounts for the $d_{xy}$ orbital:
\begin{eqnarray}
\nonumber T_{\bm k}^{11} &=& 2 t_{2} \cos k_{x} + 2 t_{1} \cos k_{y} + 4 t_{3} \cos k_{x} \cos k_{y} \\
&+& 2 t_{11} ( \cos ( 2 k_{x} ) - \cos ( 2 k_{y} ) ) \\
\nonumber &+& 4 t_{12} \cos ( 2 k_{x} ) \cos ( 2 k_{y} ) ,
\end{eqnarray}
\begin{eqnarray}
\nonumber T_{\bm k}^{22} &=& 2 t_{1} \cos k_{x} + 2 t_{2} \cos k_{y} + 4 t_{3} \cos k_{x} \cos k_{y} \\
&-& 2 t_{11} ( \cos ( 2 k_{x} ) - \cos ( 2 k_{y} ) ) \\
\nonumber &+& 4 t_{12} \cos ( 2 k_{x} ) \cos ( 2 k_{y} ) ,
\end{eqnarray}
\begin{eqnarray}
\nonumber T_{\bm k}^{33} &=& \epsilon_{0} + 2 t_{5} ( \cos k_{x} + \cos k_{y} ) + 4 t_{6} \cos k_{x} \cos k_{y} \\
&+& 2 t_{9} ( \cos ( 2 k_{x} ) + \cos ( 2 k_{y} ) ) \\
\nonumber &+& 4 t_{10} ( \cos ( 2 k_{x} ) \cos k _{y} + \cos k_{x} \cos ( 2 k_{y} ) ) ,
\end{eqnarray}
\begin{eqnarray}
T_{\bm k}^{12} &=& T_{\bm k}^{21} = 4 t_{4} \sin k_{x} \sin k_{y} ,
\end{eqnarray}
\begin{eqnarray}
T_{\bm k}^{13} &=& \bar{T}_{\bm k}^{31} = 2 i t_{7} \sin k_{x} + 4 i t_{8} \sin k_{x} \cos k_{y} ,
\end{eqnarray}
\begin{eqnarray}
T_{\bm k}^{23} &=& \bar{T}_{\bm k}^{32} = 2 i t_{7} \sin k_{y} + 4 i t_{8} \sin k_{y} \cos k_{x} .
\end{eqnarray}
In Ref.~\cite{daghofer.nicholson.12} the hopping parameters are given in electron volts as: $t_{1} = -0.08$, $t_{2} = 0.1825$, $t_{3} = 0.08375$, $t_{4} = -0.03$, $t_{5} = 0.15$, $t_{6} = 0.15$, $t_{7} = -0.12$, $t_{8} = 0.06$, $t_{9} = 0.0$, $t_{10} = -0.024$, $t_{11} = -0.01$, $t_{12} = 0.0275$ and $\epsilon_{0} = 0.75$. The average number of particles in the system $n = 4$ is attained for $\mu = 0.4748$.

The band structure of the IBSC model can be reconstructed from the kinetic tight-binding Hamiltonian in the orbital representation $H_{0}$ via the unitary transformation $H'_{0} = U^{\dagger} H_{0} U$~\cite{daghofer.nicholson.10}. Then $H'_{0} = \sum_{\varepsilon{\bm k}\sigma} E_{\varepsilon{\bm k}\sigma} d_{\varepsilon{\bm k}\sigma}^{\dagger} d_{\varepsilon{\bm k}\sigma}$. Here $d_{\varepsilon{\bm k}\sigma}$ ($d_{\varepsilon{\bm k}\sigma}^{\dagger}$) annihilates (creates) an electron in band $\varepsilon$. The total number of particles in the system $\sum_{\alpha{\bm k}\sigma} c_{\alpha{\bm k}\sigma}^{\dagger} c_{\alpha{\bm k}\sigma} = \sum_{\varepsilon{\bm k}\sigma} d_{\varepsilon{\bm k}\sigma}^{\dagger} d_{{\bm k}\varepsilon\sigma}$ is adjusted by the chemical potential $\mu$. We neglected orbital effects, which is equivalent to assuming that the external magnetic field $h$ is parallel to the FeAs layers.

For simplicity and readability we assume the existence of only the superconducting phase in the system. However, it should be had in mind that in many IBSC systems superconductivity can coexist with magnetic order.~\cite{kordyuk.12} In this paper, without  specifying the mechanisms responsible for the forming of superconducting phases, we introduce a superconducting pairing between the {\it quasi}-particles in bands $\varepsilon$, which is a good approximation in the limit of weak or vanishing inter-band pairing~\cite{hirschfeld.korshunov.11}. Superconducting states with non-zero total momentum of Cooper pairs (TMCP) can be described by the phenomenological effective Hamiltonian:
\begin{eqnarray}
H'_{SC} = \sum_{{\bm k}\varepsilon} \left( \Delta_{\varepsilon{\bm k}} d_{\varepsilon{\bm k}\uparrow}^{\dagger} d_{\varepsilon,-{\bm k}+{\bm q}_{\varepsilon}\downarrow}^{\dagger} + H.c. \right) ,
\end{eqnarray}
where $\Delta_{\varepsilon{\bm k}} = \Delta_{\varepsilon} \eta({\bm k}) = U_{\varepsilon} \eta({\bm k}) \langle d_{\varepsilon,-{\bm k}+{\bm q}_{\varepsilon}\downarrow} d_{\varepsilon{\bm k}\uparrow} \rangle$ is the SOP in band $\varepsilon$ for the TMCP ${\bm q}_{\varepsilon}$ and amplitude $\Delta_{\varepsilon}$. The structure factor given by $\eta ({\bm k})$ captures the symmetry of the SOP, related to the effective interaction in real space~\cite{ptok.crivelli.15,ptok.crivelli.13}. $H'_{SC}$ in band space is the reformulation of the interacting Hamiltonian in orbital space~\cite{daghofer.nicholson.10} Similarly to the two-band model, the SOP in the band representation can be transformed to orbital one.~\cite{ptok.14} Moreover, the interband SOPs with different values of $\Delta_{\varepsilon{\bm k}}$ in every band $\varepsilon$ correspond to the existence of the intra- and interorbital SOPs in the system.

The total Hamiltonian $H = H'_{0} + H'_{SC}$, formally describes a system with a three independent bands. For this reason, the eigenvalues of H in the band representation are given by standard Bogoliubov transformation~\cite{crivelli.ptok.14,januszewski.ptok.14}:
\begin{eqnarray}
\lambda_{\varepsilon{\bm k}}^{\pm} = \vartheta_{\bm k}^{-} \pm \sqrt{ \left( \vartheta_{\bm k}^{+} \right)^{2} + | \Delta_{\varepsilon{\bm k}} |^{2} } \quad \forall {\bm k} , 
\end{eqnarray}
where $\vartheta_{\bm k}^{\pm} = \left( E_{\varepsilon{\bm k}\uparrow} \pm E_{\varepsilon,-{\bm k}+{\bm q}_{\varepsilon}\downarrow} \right) / 2$. The grand canonical potential can be calculated {\it explicitly} from its definition $\Omega \equiv k_{B} T \ln\{\mbox{Tr}[ \exp ( - H / k_{B} T )]\}$, which for given (fixed) parameter $h$ and $T$ can be treated as function of the SOP $\Delta_{\varepsilon}$ and TMCP ${\bm q}_{\varepsilon}$ in each bands.

\subsection{The entropy and specific heat calculation}

In case of intraband superconductivity, the grand canonical potential is given as $\Omega = \sum_{\varepsilon} \Omega_{\varepsilon}$, where:
\begin{eqnarray}
\Omega_{\varepsilon} &=& -k_{B} T  \sum_{{\bm k}\alpha} \ln \left( 1 + \exp \left( - \frac{ \lambda_{\varepsilon{\bm k}}^{\alpha} }{ k_{B} T } \right) \right) \\
\nonumber &+& \sum_{\bm k} \left( E_{\varepsilon{\bm k}\downarrow} - \frac{ | \Delta_{\varepsilon{\bm k}} |^{2} }{ U_{\varepsilon} } \right) .
\end{eqnarray}
Detail calculation can be found in Ref.~\cite{crivelli.ptok.14,ptok.14,ptok.crivelli.13,januszewski.ptok.14}. From the thermodynamical potential, we can determine the entropy $S = - d\Omega / dT$ and superconducting specific heat at temperature $T$ as $C = - T \partial^{2} \Omega / \partial T^{2}$, where the $\Omega$ is the grand canonical potential. 
The entropy~\cite{wysokinski.spalek.14} is $S = - d\Omega / dT$, where:
\begin{eqnarray}
\frac{ d\Omega }{ dT } = \sum_{\varepsilon} \left[ \left( \frac{ \partial \Omega_{\varepsilon} }{ \partial T } \right)_{\mbox{e}} + \left( \frac{ \partial \Omega_{\varepsilon} }{ \partial \Delta_{\varepsilon} } \right)_{\mbox{e}} \frac{ \partial \Delta_{\varepsilon} }{ \partial T } + \left( \frac{ \partial \Omega_{\varepsilon} }{ \partial {\bm q}_{\varepsilon} } \right)_{\mbox{e}} \frac{ \partial {\bm q}_{\varepsilon} }{ \partial T } \right] 
\end{eqnarray}
where subscript $\mbox{e}$ labels the equilibrium values of the SOPs $\Delta_{\varepsilon}$ and TMCPs ${\bm q}_{\varepsilon}$. From the equilibrium condition, we have:
$\partial \Omega_{\varepsilon} / \partial \Delta_{\varepsilon} |_{\mbox{e}} = \partial \Omega_{\varepsilon} / \partial {\bm q}_{\varepsilon} |_{\mbox{e}} = 0$ for all $\varepsilon$,
since $\Omega ( \Delta_{\varepsilon} , {\bm q}_{\varepsilon} )$ is at a minimum. 
Hence $S =- \sum_{\varepsilon} \partial \Omega_{\varepsilon} / \partial T |_{\mbox{e}}$, or {\it a priori}:
\begin{eqnarray}
S = \sum_{\varepsilon{\bm k}} \left[ \frac{ \lambda_{\varepsilon{\bm k}}^{\alpha} }{ T } f ( \lambda_{\varepsilon{\bm k}}^{\alpha} ) + k_{B} \ln \left( 1 + \exp \left( - \frac{ \lambda_{\varepsilon{\bm k}}^{\alpha} }{ k_{B} T } \right) \right) \right]_{\mbox{e}}
\end{eqnarray}
where $f(E)$ is the Fermi-Dirac distribution. The specific heat is then defined in the usual manner by:
$C = T dS / dT |_{n,T,h,V} \equiv - T \partial^{2} \Omega / \partial T^{2}  |_{\mbox{e}}$.
It should be noted that the grand potential (and also SOPs and TMCPs) depend on temperature in a non-trivial manner, which forces the calculation of $S$ and $C$ to be carried out by numerical derivatives.

\begin{figure}[!b]
\begin{center}
\includegraphics{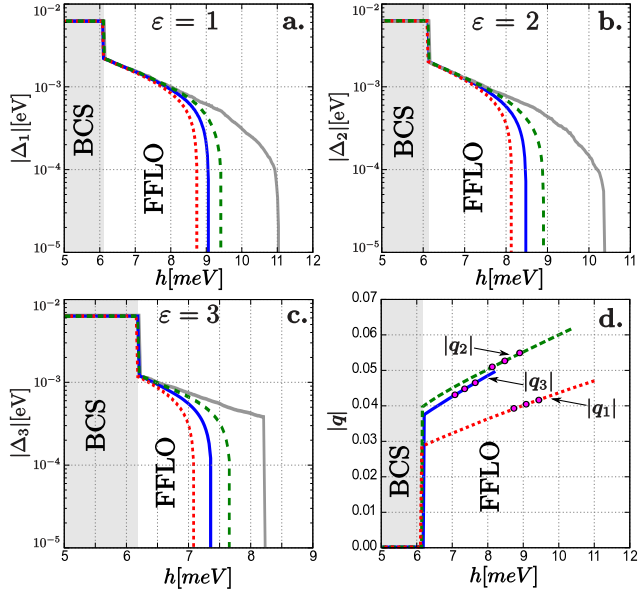}
\end{center}
\caption{(Color on-line) (a-c) Superconducting intra-band order parameter $| \Delta_{\varepsilon} |$ (energetic gap amplitude) for each band $\varepsilon$, in function of the external magnetic field for different temperatures: T equal 0.10 T$_{C}$ (red dotted line), 0.08 T$_{C}$ (blue solid line), and 0.06 T$_{C}$ (green dashed line), gray solid line denotes results for T$\simeq$0. (d) Cooper-pair total momentum ${\bm q}_{\varepsilon}$ in each band $\varepsilon$, as function of the applied magnetic field h. The lines end at the critical field of the corresponding band. The temperature dependence is negligible in the considered regime. Pink points mark the phase transitions for the different temperatures (given above).
\label{fig.1}}
\end{figure}

\section{Numerical results and discussion}

The global ground state can be obtained, at fixed values of the parameters (temperature T and magnetic field h), from the minimization of the grand canonical potential $\Omega$ with respect to the SOP $\Delta_{\varepsilon}$ and TMCP ${\bm q}_{\varepsilon}$, at the same time determining the optimal values of the latter parameters. All calculations have been performed on NVIDIA GPUs, in momentum space on a square lattice grid $k_{x} \times k_{y} = 10000 \times 10000$, using the algorithm described in Ref.~\cite{januszewski.ptok.14}. The following are predictions for \textit{s-wave} symmetry of the SOP, however other symmetries generate analogous results.

We assumed a different effective attractive intraband pairing in each band. It is found by seeking the simultaneous disappearance of the superconducting BCS phase at critical temperature T$_{C}$, equal to 5 meV ($\sim$55 K). For this set of parameters, the BCS critical magnetic h$_{C}^{BCS}$ is also uniform in the bands, with a value of about 6 meV ($\sim$103 T). Although these relatively large values can be found in some class of IBSC~\cite{gurevich.11}, we focus on generic features of IBSC in LTHM regime.

\begin{figure}[!b]
\begin{center}
\includegraphics[scale=0.9]{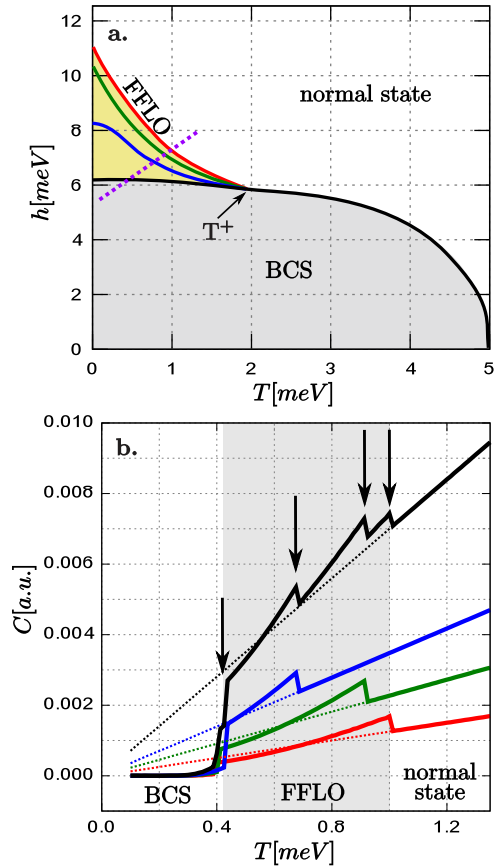}
\end{center}
\caption{(Color on-line) (a) Magnetic field $h$ -- temperature $T$ phase diagram. Solid lines denote the phase transitions between BCS state and normal or FFLO state, while red, green and blue lines phase transitions inside the FFLO state, described in the main text. (b) Specific heat along the cut line (violet dotted line in panel a). Bold black line is the total specific heat while red, green and blue solid lines are the partial specific heats for the first, second, and third band respectively. Dotted lines display the form of the specific heat for the normal state. Arrows mark the locations of phase transitions.
\label{fig.2}}
\end{figure}

Above h$_{C}^{BCS}$, at temperatures below some characteristic T$^{+}$, the FFLO phase arises~\cite{matsuda.shimahara.07}. In this phase the SOPs decrease in increasing external magnetic field (Fig.~\ref{fig.1} a-c). Moreover, in general the TMCP depend on the size of the splitting between the Fermi surfaces for electrons with spin up and down, the source of which is the Zeeman effect. Raising the external magnetic field increases the splitting, which leads to greater TMCP (Fig.~\ref{fig.1}.d)~\cite{ptok.14}, which is also true in other systems~\cite{kaczmarczyk.spalek.09}. The TMCP dependence on the temperature is however weak.

In the case of BCS, we can find a typical h--T phase diagram (Fig.~\ref{fig.2}.a), in which  superconductivity in every band disappears at the same external magnetic field~\cite{ptok.14}. Conversely for the FFLO phase, a different critical magnetic field h$_{\varepsilon, C}^{FFLO}$~\cite{ptok.crivelli.13,januszewski.ptok.14} determines the superconducting behavior in each band, as shown in Fig.~\ref{fig.2}.a. Because of this, calorimetric experiments in LTHM regime should display {\it multiple phase transitions}, in the form of a group of peaks in specific heat, shown as arrows in Fig.~\ref{fig.2}.b. The first group (or extended peak) can be connected with the transition from the BCS to the FFLO phase (first arrow from the left in Fig.~\ref{fig.2}.b), while the second group arises due to transitions inside the FFLO phase (blue and green line in Fig.~\ref{fig.2}.a), and to the final transition from superconducting to the normal state (red line in Fig.~\ref{fig.2}.a). 
To conclude, the measured total specific heat is characterized by a multiply discontinuous shape, the black line in Fig.~\ref{fig.2}.b. 
It should be noted, that similar effect can be found in systems with relatively significant finite-size effects.~\cite{aoyama.beaird.13,wojcik.zegrodnik.14,wojcik.zegrodnik.15} However, multiple phase transitions in this case (and the relative discontinuities in the specific heat), are due to abrupt changes in the TMCP ${\bm q}_{\varepsilon}$, and not the result of the disappearance of the FFLO phase in consecutive bands. This same effect can be expected in two dimensional square lattice, where a growing magnetic field favors FFLO phases with a greater number of inequivalent momenta entering the TMCP~\cite{shimahara.98}.

Usually the superconductivity in IBSC is described by an {\it s$_{\pm}$-wave} symmetry, where the gap changes its sign between the hole and electron pockets of the Fermi surface~\cite{hanaguri.niitaka.10,fuchs.drechsler.11}. However, {\it d-wave} symmetry can be also observed~\cite{maier.graser.11,abdelhafirez.grinenko.13}. Hole doping of IBSC can lead to the transition of the gap symmetry from {\it s$_{\pm}$-wave} to {\it d-wave}~\cite{tanatar.reid.10,fernandes.millis.13,watanabe.tamashita.14}. Our main result is not affected, with similar results for other symmetries than {\it s-wave} -- the FFLO is the ground state in the LTHM regime in every band. However, a different SOP symmetry could influence the shape of specific heat in function of temperature (Fig.~\ref{fig.2}). Moreover, because the band structure of the IBSC strongly depends on doping, more realistic results require specific models to the chemical compound. 

For different compounds, the quantity of specific heat peaks can depend on the number of bands forming the Fermi surface and supporting the FFLO state in the LTHM regime, with the detailed band structure influencing only quantitatively the jump heights.

\section{Summary and final remarks}

In this paper we show that the presence of the FFLO phase, in multi-band materials like IBSC, can be experimentally ascertained by the appearance of {\it multiple phase transitions}, which is in turn manifested by {\it multi-discontinuities in the shape of the specific heat}. In the context of Pauli limited superconductors, we can speak about two scales of temperature. The former is the critical temperature T$_{C}$, at which the superconductivity vanishes in zero external magnetic field. The latter is the temperature T$^{+}$, in which phase transitions from the superconducting to normal state change kind, from first to second order. T$_{C}$ is reported as 2.3 K and 3.5 K for CeCoIn$_{5}$ and KFe$_{2}$As$_{2}$ respectively, while in both cases T$^{+}$ can be approximated as 0.31 T$_{C}$~\cite{bianchi.movshovich.02,zocco.grube.13}. In CeCoIn$_{5}$ the specific heat displays an additional anomaly within the superconducting state at a temperature $\sim 300$ mK ($\sim$ 0.12 T$_{C}$)~\cite{bianchi.movshovich.03}. These experimental results are interpreted as evidence for the existence of the FFLO phase. Similar behavior is observed in organic superconductors~\cite{lortz.wang.07}.

Relevant experimental data have been presented in the literature for KFe$_{2}$As$_{2}$~\cite{kim.kim.11,abdelhafirez.aswartham.12}, but not at sufficiently low temperatures. The iron-based KFe$_{2}$As$_{2}$ Pauli-limited superconductors have critical magnetic field near 5 T~\cite{zocco.grube.13}. However also for fields above this value, an anomalous shape for the specific heat can be observed in LTHM regime (for temperatures below 1 K)~\cite{kim.kim.11,abdelhafirez.aswartham.12}. The anomalies have been related to a meta-magnetic transition in high-field heavy-fermion CeIrIn$_{5}$~\cite{takeuchi.inoue.01}, which is explained by the magnetization process in the antiferromagnet with helical spin structure. However, the heavy-fermion CeCoIn$_{5}$ {\it additional} phase transition in the LTHM regime inside the superconducting phase can be the manifestation of the emergence of incommensurate spin density wave (SDW)~\cite{kenzelmann.strassle.08}. The latter mechanism is strongly connected with the FFLO phase, because an existing SDW increases the tendency of the system toward the creation of an FFLO phase~\cite{mierzejewski.ptok.09}. To verify this hypothesis, accurate calorimetric measurements are required in LTHM regime. This phenomenon requires further experimental and theoretical studies.

\ack

The author thanks Dawid Crivelli, Przemys\l{}aw Piekarz and Andrzej \'{S}lebarski for very fruitful discussions and comments.


\end{document}